\documentclass[11 pt]{article}
\usepackage{enumerate}
\usepackage{amssymb}
\usepackage{graphicx}
\usepackage{amsthm}
\theoremstyle{plain}

\begin{document}
\title{Encoding spatial data into quantum observables}
\author{Artur Sowa \\
Department of Mathematics and Statistics\\
University of Saskatchewan \\
106 Wiggins Road,
Saskatoon, SK S7N 5E6 \\
Canada \\
sowa@math.usask.ca \\
}

\date{}
\maketitle
\newtheorem*{definition*}{Definition}
\newtheorem{theorem}{Theorem}
\newtheorem{prop}{Proposition}
\newtheorem{lemma}{Lemma}
\newtheorem{corollary}{Corollary}
\newtheorem{fact}{Fact}

 \newcommand{\sgn}{\mathop{\mathrm{sgn}}}

\begin{abstract}
 The focus of this work is a correspondence between the Hilbert space operators on one hand, and doubly periodic generalized functions on the other. The linear map that implements it, referred to as the Q-transform, enables a direct application of the classical Harmonic Analysis in a study of quantum systems. In particular, the Q-transform makes it possible to reinterpret the dynamic of a quantum observable as a (typically nonlocal) dynamic of a classical observable. From this point of view we carry out an analysis of an open quantum system whose dynamics are governed by an asymptotically harmonic Hamiltonian and compact type Lindblad operators. It is established that the initial value problem of the equivalent nonlocal but classical evolution is well posed in the appropriately chosen Sobolev spaces.  The second set of results pertains to a generalization of the basic Q-transform and highlights a certain type of asymptotic redundancy. This phenomenon, referred to as the \emph{broadband redundancy}, is a consequence of a well-known property of the zeros of the Riemann zeta function, namely, the uniform distribution modulo one of their ordinates. Its relevance to the analysis of quantum dynamics is only a special instance of its utility in harmonic analysis in general. It remains to be seen if the phenomenon is significant also in the physical sense, but it appears well-justified---in particular, by the results presented here---to pose such a question.

\vspace{.5cm}

\vspace{.5cm}

\vspace{.5cm}


\end{abstract}

\section{Introduction}

The departure point for this work is an observation that a classical doubly-periodic real function (or generalized function) can be uniquely represented as a quantum observable via an invertible linear operation dubbed the Q-transform. In particular, the time-evolution of a quantum observable, as it is described by the master equation in Lindblad form, can be translated into time-evolution of a real function. We prove that under some technical assumptions the initial value problem related to such an evolutionary process is well posed in the classical Sobolev spaces (see Theorem \ref{thA}). The proof is facilitated by two unique properties of the Q-transform. First, it furnishes the notion of Sobolev type norms on operator spaces. Moreover, such norms are submultiplicative on compact operators which refines the well known property of the Hilbert-Schmidt norm. Second, the inverse Q-transform translates the Heisenberg equation with the essentially harmonic Hamiltonian into a rudimentary first order partial differential equation, effecting a constant velocity drift of the classical alter ego of the quantum observable.

The well-posedness result may be construed as evidence for stability of classical information, be it discrete or analogue, in quantum environments\footnote{Of course, the problem of stability of discrete classical information in quantum channels is well studied, \cite{Holevo}.}. Furthermore, one can speculate that if the Q-transform were hardwired as a transducer between a classical computing machine and a carefully engineered open quantum system the resulting hybrid  might be used to effect enhancement of classical information. Indeed, numerical simulations based on methods akin to the Q-transform, \cite{Sowa33}, and more recently on the Q-transform as such, demonstrate the possibility of effecting image denoising via such a process. Note that in such an application the dissipative trends in the quantum system would be in fact beneficial which stands in contrast to the standard quantum computing schemas where they are always a detriment. The problem of designing quantum environments that effect a desired type of dissipative or non-dissipative evolution are the main focus in the area of Quantum Engineering, \cite{Zagoskin}. Separately, the Q-transform sheds some new light on the problem of simulation of quantum processes with certain properties, see \emph{Remark 1} in Subsection \ref{Main_th}.

 We point out that while there is preexisting literature pertaining to the related problem of well-posedness of the Markovian master equation, e.g. \cite{Kastor_Temme}, it appears not to be adaptable to the analysis of the dual problem considered here. Indeed, those studies are set in the context of noncommutative $\mathbb{L}_p$ spaces and rely heavily on the probabilistic interpretation of the density matrix and the analogies with the classical Markov processes that it brings out. Separately, in recent times there has been a lot of interest in the related topic of evolution equations with a nonlocal term that is either linear, e.g. \cite{Caffarelli}, or nonlinear, e.g. \cite{Himonas}. The dominant methodology of those studies is that of geometric analysis as well as the geometric theory of Banach spaces, which stands in contrast to the operator-algebraic arguments used in the present work.

In the second part of this article we discuss the notion of generalized Q-transforms which are obtained by passing from the Fourier basis to other, generically non-orthonormal, bases of $L_2[0, 1]$. This creates the right framework for a discussion of the newly discovered phenomenon of \emph{broadband redundancy}, \cite{Sowa34}. The main findings are summarized in Theorems \ref{theorem_special_D} and \ref{thB}. The former theorem illustrates how the broadband redundancy is manifested in the classical harmonic analysis of periodic single-variable functions. The latter shows how it manifests itself in two dimensions and how that translates into the language of quantum theory. It is likely that neither one of these two theorems is optimal in its present form as, indeed, based on conjectures discussed in \cite{Sowa34} one may expect stronger statements to be true. Finally, we point out that at this stage too little is known to be sure whether the broadband redundancy is only a mathematical concept applicable to engineered systems or, indeed, a phenomenon that occurs in nature which ought to be considered in the fundamental quantum theory. Further analysis will hopefully lead to conceptualization of physical hypotheses and experimental tests to address this problem.

\section{The concept of the Q-transform}

In what follows we use a few core properties of the Sobolev spaces on the two-dimensional torus.  It is of importance with regards to our purposes that the torus is rather special in that its Sobolev spaces are directly tied to the Fourier series. The very basic treatment of Sobolev spaces on a circle may be found in \cite{Warner} or \cite{Kress}, and the generalization to higher-dimensional tori is essentially trivial. Also, an interesting tori-specialized proof of the Sobolev inequality is given in \cite{Benyi-Oh}\footnote{It highlights the remarkable idiosyncrasies of this case as compared to that of general manifolds, e.g. as in \cite{Aubin}.}.

Let $\mathbb{T}= \mathbb{R}^2/\mathbb{Z}^2$ be the standard two-dimensional torus. We will denote generic real functions $f, g, h:\mathbb{T} \rightarrow \mathbb{R}$, and complex valued functions $u, v, w:\mathbb{T} \rightarrow \mathbb{C}$.  When convenient we will identify functions on a torus with doubly periodic functions on the $\mathbb{R}^2$ plane without a comment.
Next, recall that for an arbitrary quantum system its observables are synonymous with the Hermitian operators on a specific, system-determined, Hilbert space. The operators need not be bounded, e.g. the energy observable (the Hamiltonian) is frequently unbounded. We will denote the generic observables $\mathcal{A}, \mathcal{B}$, etc. They satisfy $\mathcal{A}^\dagger = \mathcal{A}$ where the Hermitian adjoint $\dagger$ is defined via the Hilbert structure in the usual way. Also, general operators will be denoted $A,B,C $, etc. Recall  that any non-Hermitian operator can be uniquely represented as the sum of a Hermitian operator and an anti-Hermitian one, i.e.
\begin{equation}\label{H_antiH}
   C= \mathcal{A} + i \mathcal{B} \quad \mbox{ where } \mathcal{A} = \frac{1}{2}(C + C^\dagger) \,\, \mbox{ and  } \mathcal{B} =\frac{1}{2i}(C - C^\dagger).
\end{equation}
It is shown below that there exists a natural correspondence between distributions on $\mathbb{T}$ and operators on a Hilbert space, in which real-valued distributions correspond to observables. Formally, if $u = f+ig$  we set
\begin{equation}\label{Q}
Qf = \mathcal{A},\,\, Qg = \mathcal{B},  \quad \mbox{ and } \quad Qu = Qf+iQg = \mathcal{A} + i \mathcal{B},
\end{equation}
we will refer to it as the Q-transform.
This furnishes a method of passing (or transducing) classical information onto quantum observables. Vice-versa, in this way one can visualize quantum observables via graphs of real functions. It also leads to the notion of Sobolev norms on operator spaces, a concept that enables control of the regularity of classical functions corresponding to quantum observables, and is foundational in the study of well-posedness of a quantum process.

\subsection{Real functions on a 2D-torus compared to Hermitian operators}\label{subsection-define-S}

Let $f$ be real and integrable, i.e. $f\in L_1(\mathbb{T})$, the Fourier coefficients of $f$ are defined via
\[
z_{k,l} =  \hat{f}_{k,l}  =  \int\limits_0^{1}\int\limits_0^{1} f(x,y) e^{-2\pi i (kx+ly)} \, dxdy.
\]
Since $f$ is real-valued the infnite matrix $\hat{f} = Z = [z_{k,l}]_{(k,l)\in\mathbb{Z}^2} $ has the following symmetry:
\begin{equation}\label{point-sym}
z_{-k,-l} = {z_{k,l}}^*,
\end{equation}
where $^*$ denotes the complex conjugate. Next, denote $x_{k,l} = \Re \, z_{k,l}$, $y_{k,l} = \Im\, z_{k,l}$, and let $W = [w_{k,l}]_{(k,l)\in\mathbb{Z}^2} $ be defined by setting
\[
w_{k,l} = z_{k,l} \quad \mbox{ whenever } k<l,
\]
\[
w_{k,l} = {z_{l,k}}^* \quad \mbox{ whenever } k> l,
\]
\[
w_{k,k} = \left\{\begin{array}{cc}
            \sqrt{2}\, y_{k,k}, & k< 0 \\
            & \\
            \sqrt{2}\, x_{k,k}, & k>0
          \end{array}\right.
          \quad \mbox{ and }\quad  w_{0,0}= z_{0,0} \in \mathbb{R}.
\]
Observe that $W$ is a Hermitian matrix, i.e. $W^\dagger = W$, where $\dagger$ denotes  complex conjugation followed by matrix transposition. We will denote this symmetry changing construction by the letter $S$ and write $W = S[Z]$. Note that $S$ is invertible, i.e. a Hermitian matrix $W$ gives rise to a matrix $Z = S^{-1}[W]$ that satisfies symmetry (\ref{point-sym}). The construction of $Z= S^{-1}[W]$ consists of copying the above-diagonal part of $W$ into $Z$, extending $Z$ below the diagonal via (\ref{point-sym}), and finally setting the diagonal terms according to
\[
z_{k,k} = \left\{\begin{array}{cc}
            (w_{-k,-k} + i w_{k,k})/\sqrt{2}, & k< 0 \\
            & \\
            (w_{k,k} - i w_{-k,-k})/\sqrt{2}, & k>0
          \end{array}\right.
          \quad \mbox{ and }\quad  z_{0,0}= w_{0,0} \in \mathbb{R}.
\]

\vspace{.5cm}

\begin{figure}[ht!]
\includegraphics[width=160mm]{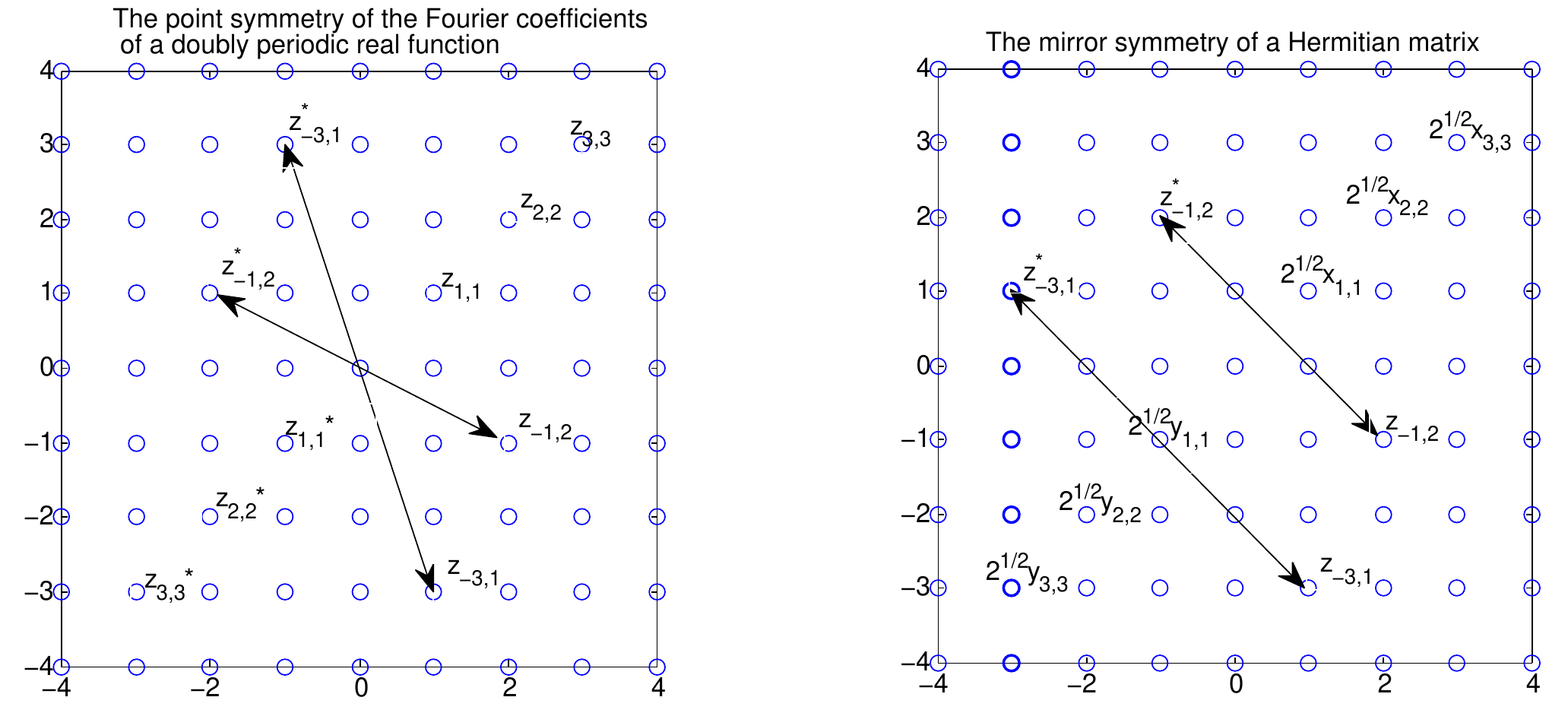}
\caption{The basic symmetry underlying the $Q$-transform. The Fourier coefficients of a doubly periodic real function satisfy the symmetries indicated schematically in the left figure whereas a rearrangement of the same data in accordance with the schema displayed in the right figure yields a Hermitian matrix. Note that the order of the vertical (i.e. first) indices is unconventional for a matrix, hence the diagonal runs from the lower left to the upper right. }
\label{quantumflow}
\end{figure}

\noindent The following observation will be indispensable in what follows.
\begin{lemma}\label{rot_sym_lem}
For any $\gamma: \mathbb{Z}^2 \rightarrow \mathbb{R}$ rotationally symmetric and non-negative, i.e. $\gamma(k,l) = \tilde\gamma (k^2+l^2)\geq 0$,  and $W=S[Z]$ we have
\begin{equation}\label{rot_sym}
\sum\limits_{(k,l)\in\mathbb{Z}^2} \gamma(k,l) \, |w_{k,l}|^2 = \sum\limits_{(k,l)\in\mathbb{Z}^2} \gamma(k,l) \, |z_{k,l}|^2,
\end{equation}
i.e. the two sides are either simultaneously infinite or  simultaneously finite and equal.
\end{lemma}
\vspace{.2cm}

\noindent
{\em Proof.}
This identity follows directly from the very construction of $W$ from $Z$ (see Figure \ref{quantumflow}). $\Box$
\vspace{.5cm}

Let us now fix a Hilbert space, say, $\mathbb{H}$ with a distinguished orthonormal basis, say, $(e_n)$ where $n\in \mathbb{Z}$. A Hermitian matrix $W$ determines an operator (also denoted $W$) by setting $W e_n = \sum_{k}w_{k,n}e_k$, and extending the action of $W$ to other vectors by linearity. It is clear that the matrix identity $W^\dagger = W$ implies that $W$ is a self-adjoint operator. Its domain consists of those $x\in\mathbb{H}$ for which $Wx\in\mathbb{H}$.

Similarly, a matrix $Z$ that displays symmetry (\ref{point-sym}) can often be interpreted as the Fourier transform of a distribution. Recall that a periodic distribution is a continuous linear functional  $\omega: C^\infty(\mathbb{T}) \rightarrow \mathbb{R}$, where $C^\infty(\mathbb{T})$ is the space of smooth real functions on a torus endowed with the usual $C^\infty$ topology. The Fourier coefficients of $\omega$ are defined via $\hat{\omega}_{k,l} = \omega(e^{-2\pi i (kx+ly)})$. Thus, a matrix $Z$ may in some instances determine a distribution $\omega$ by requiring $\hat{\omega}_{k,l}= z_{k,l}$ or, equivalently, by defining $\omega$ directly as follows:
\[
\mbox{ For }  f = \sum\limits_{k,l} \hat{f}_{k,l}e^{-2\pi i (kx+ly)} \quad \mbox{ define }\quad
\omega_Z(f) = \sum\limits_{k,l} \hat{f}_{k,l}z_{k,l}.
\]
This suggests that the symmetry change operation $S$ furnishes a bridge between Hilbert space operators on one side and periodic distributions on the other.
However, a possible invertibility of this operation in the rigorous analytical sense is still in question as, indeed,  it is not \emph{a priori} clear what conditions on $Z$ ensure continuity of the corresponding functional $\omega_Z$. In the next section we will dispel this problem  by focusing on the Sobolev classes of distributions. With this understood we introduce a purely formal definition of the Q-transform. Its meaning is made rigorous in Proposition \ref{Q_precise}.

\begin{definition*}
Let $f, g:\mathbb{T} \rightarrow \mathbb{R}$ be integrable functions or periodic distributions, and $u = f+ig$. We define the Q-transform of $f$ to be the observable $Qf = S[{\hat{f}}]$ and the Q-transform of $u$ to be the operator $Qu = Qf +iQg$. Conversely, let $C$ be an arbitrary operator and let $ C= \mathcal{A} + i \mathcal{B}$ be its representation as the sum of a Hermitian and anti-Hermitian parts. We define its Q-inverse to be $u = Q^{-1}\mathcal{A} + iQ^{-1}\mathcal{B} = (S^{-1}\mathcal{A})^\vee + i(S^{-1}\mathcal{B})^\vee $.
\end{definition*}
\vspace{.1cm}

\noindent
\textbf{Remark 1.} We emphasize that the $Q$ transform depends on the choice of basis in the space $\mathbb{H}$. Equivalently, one could choose any one from among the family of $U^\dagger Qu U$ where $U$ runs over the set of unitary operators on $\mathbb{H}$.
\vspace{.1cm}

\noindent
\textbf{Remark 2.} Note that we obtain a variant of the Q-transform by restricting attention to a finite-dimensional Hilbert space $\mathbb{H}$ with basis $(e_n)$ and $n\in \{-N, -N+1, \ldots, -1, 0 , 1, \ldots , N-1, N\}$ on one hand, and the space of trigonometric polynomials of degree not exceeding $N$ on the other. The above definition of the Q-transform is fully rigorous in the finite-dimensional case, and requires no further justification. This in essence is how the Q-transform is implemented in numerical simulations.

\subsection{Norms invariant under the Q-transform}

For  $u,v:\mathbb{T} \rightarrow \mathbb{C}$ and real $\alpha \geq 0$ we have the Sobolev sesquilinear form
\[
\langle u\,| \,v\rangle_{\alpha} = \sum\limits_{(k,l)\in\mathbb{Z}^2} (1+k^2 +l^2)^{\alpha} \, \hat{u}_{k,l}\,{\hat{v}_{k,l}}^*
\]
as well as the Sobolev norm
$
\| u\|_{\alpha} = \sqrt{\langle u\,| \,u\rangle_{\alpha} }
$. Note that for $u=f+ig$ we have
\begin{equation}\label{Pyth_Sob}
  \| u\|_{\alpha}^2 = \| f\|_{\alpha}^2 +  \|g\|_{\alpha}^2.
\end{equation}
As is typical we let $H^\alpha_C$  denote the set of equivalence classes of measurable functions (with two functions being equivalent if they differ on a set of Lebesgue measure zero in $\mathbb{T}$) with the finite Sobolev $\alpha$-norm. It is well known that $H^\alpha_C$ with the  scalar product $\langle\,\, |\,\,\rangle _{\alpha}$ is a separable Hilbert space. In particular, $H^0_C = L_2(\mathbb{T})$. Recall, \cite{Aubin}, that the greater $\alpha$ the higher the regularity of every $u\in H^\alpha_C$. In particular, by virtue of the Sobolev lemma, $u$ is continuous whenever $\alpha > 1$. More generally, if $\alpha > 1+k$, then $u\in H^\alpha_C$ implies $u\in C^{k}(\mathbb{T})$, and $\|u\|_{C^k}\leq C \|u\|_\alpha$ for a constant $C$ independent of $u$. Separately, the Rellich lemma, \cite{Warner}, states that $H^\alpha_C\subset H^\beta_C$ whenever $\alpha > \beta$, and the embedding is a compact operator.

Moreover, as is common  we define $H^{-\alpha}_C$ for $\alpha \geq 0$ to be the dual space of $H^{\alpha}_C$. For a linear functional $\omega: H^{\alpha}_C \rightarrow \mathbb{C}$ the Fourier coefficients are defined, as above, $\hat{\omega}_{k,l} = \omega(e^{-2\pi i (kx+ly)})$.
It is well known and easy to verify that $H^{-\alpha}_C$ consists precisely of those linear functionals on $H^{\alpha}_C$ that have a finite $(-\alpha)$-norm, i.e.
\[
H^{-\alpha}_C = \left\{ \omega: H^{\alpha}_C \rightarrow \mathbb{C}: \sum\limits_{(k,l)\in\mathbb{Z}^2} (1+k^2 +l^2)^{-\alpha} \, |\hat{\omega}_{k,l}|^2 < \infty \right\}.
\]
In particular, $H^{-\beta}_C\subset H^{-\alpha}_C$ whenever $\alpha > \beta$. It is also well known that the set $H^{-\infty}_C = \bigcup_{\alpha > 0}H^{-\alpha}_C$ is the space of all periodic distributions.

Next, we switch focus to operators and introduce a generalization of the Hilbert-Schmidt norm for operators. We begin by distinguishing an orthonormal basis $(e_n)_{n\in\mathbb{Z}}$ in $\mathbb{H}$. It is well known, \cite{Akhiezer}, that every bounded operator $A: \mathbb{H}\rightarrow\mathbb{H}$ admits a matrix representation, which is given by matrix entries $a_{k,l} = \langle e_k| Ae_l \rangle$. When an operator is unbounded the matrix needs not be well defined as, indeed, even $Ae_l$ may be undefined for some $l$. However, it is well known that if $A$ is densely defined, then there exists an orthonormal basis whose elements belong to its domain. Therefore a matrix may be defined even for a densely-defined unbounded operator provided the distinguished basis is carefully chosen. In what follows we define classes of operators via specific properties of their matrix representation. We then prove some of their properties, the purpose of which is to enable analysis of classical information encoded in quantum observables. When the results are applied in Section \ref{section_dynamics}, we are careful to chose the distinguished bases so that the relevant operators admit matrix representations.

With this understood, for $A = [a_{k,l}]$ and $B = [b_{k,l}]$ as well as $\alpha \in \mathbb{R}$ (positive or not) we set
\[
\langle A\,| \,B\rangle_{\alpha} = \sum\limits_{(k,l)\in\mathbb{Z}^2} (1+k^2 +l^2)^{\alpha} \, a_{k,l}\,b_{k,l}^* ,
\]
as well as $\| A\|_{\alpha} = \sqrt{\langle A\,| \,A\rangle_{\alpha} } $.
We define Sobolev spaces of operators by requiring finiteness of the relevant Sobolev norms, i.e.
\[
H^\alpha_Q = \{A: \mathbb{H}\rightarrow\mathbb{H}: \| A\|_{\alpha} < \infty\quad\mbox{ w.r.t. the distinguished basis } (e_n)_{n\in\mathbb{Z}} \}.
 \]
In particular, $A\in H^\alpha_Q$ implies that $A$ has a well defined matrix with respect to basis $(e_n)$. This is a restrictive condition only in the case of $\alpha <0$. Indeed, if $\alpha \geq0$, then $A$ is automatically Hilbert-Schmidt, hence bounded (even compact) so that $A$ has a well defined matrix regardless of the choice of $(e_n)$. Note that for $C=\mathcal{A}+i\mathcal{B}$ we have
\begin{equation}\label{Pyth_Sob-Noncom}
  \| C\|_{\alpha}^2 = \| \mathcal{A}\|_{\alpha}^2 +  \|\mathcal{B}\|_{\alpha}^2.
\end{equation}
Next, we wish to demonstrate that the spaces $H^\alpha_Q $ are in fact Hilbert spaces. However, this is straightforward:
\begin{prop} \label{Q_precise}
Fix $\alpha \in \mathbb{R}$. Then $ Q : H^\alpha_C \rightarrow H^\alpha_Q$ is a norm preserving (complex-)linear bijection. In particular, $H^\alpha_Q$ with the sesquilinear form $\langle\,\,|\,\,\rangle_\alpha$ is a separable Hilbert space.
\end{prop}
\vspace{.2cm}
\noindent
{\emph Proof.} First, clearly, for $u=f+ig$ and $z=x+iy$ one has  $Q(zu) =zQ(u)$, so that $Q$ is complex linear. Thus, the statement is a direct consequence of (\ref{Pyth_Sob}), (\ref{Pyth_Sob-Noncom}), and Lemma \ref{rot_sym_lem}.  $\Box$
\vspace{.2cm}

\noindent
\textbf{Remark 1.} Henceforth we will use the Q-transform only as acting between Sobolev spaces $ Q : H^\alpha_C \rightarrow H^\alpha_Q$ with fininte $\alpha$ or, if needed, in the finite-dimensional setting. This gives a rigorous meaning to the informal definition of the Q-transform given in Subsection \ref{subsection-define-S}.
\vspace{.1cm}

\noindent
\textbf{Remark 2.} Note that the definition of $H^\alpha_C$ involves equivalence classes of functions, while there is no need for anything of the sort in the definition of $H^\alpha_Q$.  That, of course, stems from the fact that in the latter case there is no need for the Fourier transform or its many intricacies inherited from the measure theory.
\vspace{.1cm}

\noindent\textbf{Remark 3.}
We note that the $H^0_Q$ norm is identical with the well-known Hilbert-Schmidt norm, i.e.
\[
\|A\|_{0}^2 = \sum\limits_{(k,l)\in\mathbb{Z}^2} |a_{k,l}|^2 = \mbox{ trace }AA^\dagger.
\]
In particular, by cyclicity of the trace we have $\|U^\dagger AU\|_{0} = \|A\|_{0}$ for any unitary operator $U$.
When $\alpha\neq 0$ the $\alpha$-norm is not a unitary invariant and $\|U^\dagger \mathcal{A}U\|_{\alpha}$ generally depends on the unitary matrix $U$. However, most importantly,
\begin{equation}\label{invariance}
  \|U^\dagger AU\|_{\alpha} = \|A\|_{\alpha} \quad \mbox{ whenever } U \mbox{ diagonal in  } (e_n).
\end{equation}
Indeed, this follows from an observation that the matrix entries of $U^\dagger AU$ differ from those of $A$ only by phase factors. We also note the simple property $\|A\|_\alpha=\|A^\dagger\|_\alpha$.

\subsection{Further properties of the Sobolev norms of operators}\label{subsection_sub_mult}

\noindent
As it turns out, positive Sobolev norms ($\alpha\geq 0$) of operators are sub-multiplicative. Namely, we have the following

\begin{prop} \label{sub_mult}
Assume that $A, B\in H^\alpha_Q$ (not necessarily Hermitian) with a fixed but arbitrary $\alpha \geq 0$. Then $AB\in H^\alpha_Q$ and, moreover,
\[
\|AB\|_{\alpha} \leq  \|A\|_{\alpha}\|B\|_{\alpha}.
\]
In particular, $H^\alpha_Q$ with the standard commutator $[A,B] = AB-BA$ is a Lie algebra.
\end{prop}
\vspace{.5cm}

\noindent
{\em Proof.} We will use the trivial inequality
\[
(1+j^2+k^2)^\alpha  \leq (1+j^2+p^2)^\alpha(1+r^2+k^2)^\alpha \quad (\alpha \geq 0,\,\, j,k,p,r \in \mathbb{Z}).
\]
We have,
\[\begin{array}{cl}
\|A\|^2_{\alpha}\,\|B\|^2_{\alpha} &  = \sum\limits_{j,p} |a_{j,p}|^2(1+j^2+p^2)^\alpha\sum\limits_{r,k} |b_{r,k}|^2(1+r^2+k^2)^\alpha\\
     &  \\
     & =  \sum\limits_{j,k}\sum\limits_{r,p} |a_{j,p}|^2 |b_{r,k}|^2(1+j^2+p^2)^\alpha(1+r^2+k^2)^\alpha\\
     & \\
     & \geq  \sum\limits_{j,k}(1+j^2+k^2)^\alpha\sum\limits_{r,p} |a_{j,p}|^2 |b_{r,k}|^2 \\
     & \\
     & = \sum\limits_{j,k}(1+j^2+k^2)^\alpha\sum\limits_{p} |a_{j,p}|^2 \sum\limits_{r}|b_{r,k}|^2\\
     & \\
     & \geq \sum\limits_{j,k}(1+j^2+k^2)^\alpha |\sum\limits_{p} a_{j,p} b_{p,k}|^2 = \|AB\|^2_{\alpha}
    \end{array}
\]
In particular $\|[A,B]\|_\alpha \leq 2\|A\|_\alpha\|B\|_\alpha$, which shows that $H^\alpha_Q$ is closed with respect to the bracket operation, and therefore is a Lie algebra. This completes the proof. $\Box$
\vspace{.2cm}

In particular, Proposition \ref{sub_mult} implies that if $A, B\in H^\alpha_Q$ for $\alpha \geq 0$, then $[A, B]\in H^\alpha_Q$. However, the assumption $\alpha \geq 0$ is essential. Separately, we have the following:

\begin{prop}
Let $\mathcal{H}, \mathcal{A}$ be arbitrary Hermitian operators, and let $\alpha\in \mathbb{R}$ be arbitrary. If $\langle [\mathcal{H}, \mathcal{A}]\,| \,\mathcal{A}\rangle_{\alpha}$ is finite, then it is purely imaginary, i.e.
$
\Re \,\langle [\mathcal{H}, \mathcal{A}]\,| \,\mathcal{A}\rangle_{\alpha} =0 .
$
\end{prop}
\vspace{.1cm}

\noindent
{\em Proof.} Let $\mathcal{H} = [h_{k,l}]$,
$\mathcal{A}= [a_{k,l}$], where both matrices are of the same finite size or both are infinite. Let us denote
\[
C_{r,k} := \sum\limits_{l} (1+k^2+l^2)^\alpha \, a_{r,l}a_{k,l}^*.
\]
It is seen by inspection that
\[
\sum\limits_{k} (1+k^2+l^2)^\alpha \, a_{k,r}a_{k,l}^* = C_{r,l}^*.
\]
We use this fact as follows:
\[
\begin{array}{rl}
  \langle [\mathcal{H}, \mathcal{A}]\,| \,\mathcal{A}\rangle_{\alpha} = & \sum\limits_{k,l} (1+k^2+l^2)^\alpha \,
  \sum\limits_{r} (h_{k,r} a_{r,l}a_{k,l}^* -   a_{k,r}h_{r,l}a_{k,l}^* )\\
  & \\
  = & \sum\limits_{k}\sum\limits_{r} h_{k,r} C_{r,k} -  \sum\limits_{l}\sum\limits_{r} h_{r,l} C_{r,l}^* =
  \sum\limits_{k}\sum\limits_{r} 2i\, \Im ( h_{k,r}C_{r,k} ),
\end{array}
\]
i.e. $\langle [\mathcal{H}, \mathcal{A}]\,| \,\mathcal{A}\rangle_{\alpha}$ is purely imaginary, if finite, as claimed. $\Box$

\subsection{Further properties of the Q-transform}

The Q-transform leads to a definition of a nonstandard product of any complex functions on the torus as well as a well defined commutator for any real functions or even distributions on $\mathbb{T}^2$. Namely,
\begin{definition*}
For  $f,g\in H^\alpha_C$ with $\alpha \in \mathbb{R}$, let
\begin{equation}\label{commutator}
   [f,g] := Q^{-1} i[Qf,Qg],
\end{equation}
where $[Qf,Qg] = QfQg-QgQf$ is the standard commutator of two operators. Since $i[Qf,Qg]$ is Hermitian, the commutator $[f,g]$ is real. The definition of the commutator is then extended to complex valued $u,v\in H^\alpha_C$ by requiring bi-linearity.
\end{definition*}
We have the following

\begin{prop}\label{Lie}
$H^\alpha_C$ with $\alpha \geq 0$ with the commutator defined above is a complex Lie algebra. \emph{A fortiori}, the subset of $H^\alpha_C$ that consists of real valued functions, denoted $\Re H^\alpha_C$, with the same commutator is a real Lie algebra.
\end{prop}
\vspace{.2cm}

\noindent
{\em Proof.} First, it follows from Propositions \ref{Q_precise} and \ref{sub_mult} that if $u,v\in H^\alpha_C$, then
\[
\|[u,v]\|_\alpha = \|i[Qu,Qv]\|_\alpha = \|[Qu,Qv]\|_\alpha \leq 2\|Qu\|_\alpha\|Qv\|_\alpha=2\|u\|_\alpha\|v\|_\alpha,
\]
i.e. $[u,v] \in H^\alpha_C$, so $H^\alpha_C$ is closed with respect to the commutator. Second, it is clear that the commutator is bilinear and antisymmetric. Furthermore, the Jacobi identity follows directly from (\ref{commutator}).
This completes the proof. $\Box$

\section{An application to quantum dynamics}
\label{section_dynamics}

Consider evolution of quantum observables in the Heisenberg picture, which is expressed via the master equation in Lindblad form,  \cite{Alicki}, namely:
\begin{equation}\label{Lindblad_master_dual}
\partial_t\mathcal{A} = i [\mathcal{H},\mathcal{A}] + \sum\limits_j \{L_j^\dagger \mathcal{A} L_j - \frac{1}{2} L_j^\dagger L_j\mathcal{A} - \frac{1}{2} \mathcal{A} L_j^\dagger L_j \}
\end{equation}
 Recall that the Hamiltonian $\mathcal{H}: \mathbb{H}\rightarrow\mathbb{H}$ is Hermitian while neither one of the finite collection of operators $L_j: \mathbb{H}\rightarrow\mathbb{H}$ need be Hermitian. The basic theory concerning the master equation is already classical, \cite{Alicki}, \cite{Pettrucione}, \cite{Percival}, although many sophisticated questions remain  and continue to inspire contemporary research, e.g. \cite{Breuer}, \cite{Kastor_Temme}. Most results focus on the Schr\"{o}dinger picture, wherein it is the state of the open quantum system that evolves while observables remain frozen in time. The Schr\"{o}dinger picture lends itself to the preexisting methods of analysis more easily than (\ref{Lindblad_master_dual}) which is partly due to the fact that its structure is analogous to that of the classical Markov processes.

Our focus with regards to (\ref{Lindblad_master_dual}) is on the translation of the dynamic of $\mathcal{A}$ into the dynamic of a real function which is obtained by setting $f_t = Q^{-1}\mathcal{A}(t)$. It is natural to ask if $f_t$ preserves its regularity, i.e. if one might expect that $f_0\in H^\alpha$ ensures $f_t\in H^\alpha$ for all $t\geq 0$ and if so, whether $f_t$ depends continuously (in the $\alpha$-norm) on the initial value $f_0$. Theorem \ref{thA} gives an affirmative answer to both questions under certain technical assumptions. We build a proof of this result in steps by first considering the non-dissipative and the pure Lindblad cases and then addressing the more general case via an argument based on the Trotter formula.

\subsection{The non-dissipative case}\label{subsection_non_dissip}
The first set of questions pertains to the non-dissipative case, i.e. an isolated quantum system. In such a case (\ref{Lindblad_master_dual}) simplifies to the Heisenberg equation
 \begin{equation}\label{Heisenberg}
\partial_t\mathcal{A} = i [\mathcal{H},\mathcal{A}],
\end{equation}
whose solution is determined by the initial condition via
 \begin{equation}\label{Heisenberg_sol}
\mathcal{A}(t) = \exp(i\mathcal{H}t) \mathcal{A}(0) \exp(-i\mathcal{H}t).
\end{equation}
 We will interpret this dynamic in the framework of evolution of real functions on a torus. Let the Hamiltonian be of the form
  \begin{equation}\label{H_harmon}
 \mathcal{H}= \sum\limits_{n\in \mathbb{Z}} h_n |e_n\rangle\langle e_n|: \mathbb{H} \rightarrow \mathbb{H},
  \end{equation}
where $(e_n)_{n\in\mathbb{Z}}$ is an orthonormal basis in $\mathbb{H}$. For simplicity we choose $(e_n)$ to be the distinguished basis defining the Q-transform.  Furthermore, let $[a_{k,l}(t)]$ be the matrix coefficients of $\mathcal{A}(t)$ in the basis $(e_n)$ so that (\ref{Heisenberg_sol}) is equivalent to
\begin{equation}\label{the_as}
a_{k,l} (t) = a_{k,l} (0) e^{i(h_k-h_l)t}.
\end{equation}
Next, defining $f_t = Q^{-1} \mathcal{A}(t)$ we have
\begin{equation}\label{Heisenberg_f}
\begin{array}{rl}
  f_t(x,y) =   a_{0,0}(0)\,\,  +& \sqrt{2}\sum\limits_{k >0}( a_{k,k} (0) \cos{(2\pi  k(x +y))}+ a_{-k,-k} (0) \sin{(2\pi  k(x +y))}) \\
   &  \\
     + & 2\Re\, \sum\limits_{k<l} a_{k,l} (0) e^{i(h_k-h_l)t + 2\pi i (kx + ly)}.
\end{array}
\end{equation}
 Summarizing, we obtain the following

\begin{prop} \label{prop_Harmonic}
Assume $h_n = an + c$ for some real constants $a$ and $c$. Let $\mathcal{A}(t)$ be a solution of (\ref{Heisenberg}) with $\mathcal{A}(0)\in H^\alpha_Q$ with $\alpha \in \mathbb{R}$. Then $\mathcal{A}(t)\in H^\alpha_Q$ for all times $t$. Also $f_t = Q^{-1}\mathcal{A}(t)\in H^\alpha_C$ for all $t$ and, moreover,
\begin{equation}\label{Harmonic_sol}
f_t(x,y) = f_0(x+v_xt, y+v_yt) \quad \mbox{ where }\, v_x = -\frac{a}{2\pi}, v_y = \frac{a}{2\pi}.
\end{equation}
\end{prop}
\vspace{.2cm}

\noindent
{\em Proof.} The first statement follows from (\ref{invariance}) and, by Proposition \ref{Q_precise}, $f_t$ is in $H^\alpha_C$. Furthermore, (\ref{Harmonic_sol}) is obtained by substituting $h_k-h_l = a(k-l)$ in (\ref{Heisenberg_f}).  $\Box$
\vspace{.5cm}

It follows from  the Sobolev embedding theorem that when $\alpha$ is sufficiently large (the threshold value is not important here) $f(t,x,y)=f_t(x,y)$ as above is differentiable with respect to $(x,y)$. In such a case (\ref{Harmonic_sol}) implies that $f$ is also differentiable with respect to $t$ and  satisfies a first order partial differential equation:
\begin{equation}\label{local_DE}
\partial_t \, f = \frac{a}{2\pi}(\partial_y - \partial_x) \,f.
\end{equation}
Therefore, quite remarkably, the Q-transform establishes a notion of equivalence between the evolution driven by a harmonic oscillator with the time evolution governed by a completely \emph{local} law (\ref{local_DE}). We emphasize that this property is retained in finite dimensions (cf. Remark 2 in Subsection \ref{subsection-define-S}) in which case it is an applicable result. In the case of a general Hamiltonian this type of behaviour cannot be expected. Indeed, the different pure modes of $f_t$ (i.e. its components corresponding to spacial frequencies $(k,l)$) will be moving with different velocities. It is easily seen by means of numerical experimentation that the mixing of various modes resulting form such a process can all but erase any perceivable characteristic of $f_0$ albeit, of course, the process is always reversible via a change of the time direction.
\vspace{.2cm}

\noindent
\textbf{Remark.} The eigenvalues of Hamiltonian (\ref{H_harmon}) are not bounded below. In applications one may wish to consider a modified Hamiltonian that is bounded below. Let us choose $n_0<0$ and set $ h_n =0$ for all $n<n_0$ while $h_n = an+c$ for all $n\geq n_0$. Next, assume $a_{k,l}(0) = 0$ whenever $k< n_0$ or $l<n_0$.  Then, Proposition \ref{prop_Harmonic} holds without change. Indeed, in light of (\ref{the_as}) we have $a_{k,l}(t)=0$ for all $t>0$ whenever $a_{k,l}(0)=0$. Hence, formula (\ref{Heisenberg_f}) and the proof of the proposition are unaffected by this specific modification of $\mathcal{H}$. In particular, the result applies to the finite-dimensional harmonic Hamiltonian.

\subsection{Dynamics with a compact Hamiltonian and dissipation}
Next, we consider the special case of (\ref{Lindblad_master_dual}) in which the Hamiltonian $\mathcal{H} = \mathcal{C}\in H^\alpha_Q$ for some $\alpha \geq 0$, i.e. in particular $\mathcal{C}$ is a compact operator. This includes the case when $\mathcal{H} =0$, i.e. the pure Lindblad flow. (Note that the same type of dynamics occurs also when the Hamiltonian is fully degenerate, i.e. possesses only one energy level, so that $[\mathcal{H},\mathcal{A}]=0$.) Namely, let
\begin{equation}\label{Lindblad_pure}
\partial_t\mathcal{A} = i\mathcal{C}\mathcal{A} -i \mathcal{A}\mathcal{C} + \sum\limits_{j=1}^J \{L_j^\dagger \mathcal{A} L_j - \frac{1}{2} L_j^\dagger L_j\mathcal{A} - \frac{1}{2} \mathcal{A} L_j^\dagger L_j \}
\end{equation}
 We make the following observation

\begin{prop} \label{prop_Lind_gen}
Let $\mathcal{A}(t)$ satisfy (\ref{Lindblad_pure}) with $\mathcal{C}\in H^\alpha_Q$ and $L_j\in H^\alpha_Q$ for $j = 1,2,\ldots, J$. If $\mathcal{A}(0)\in H^\alpha_Q$ ($\alpha \geq 0$), then $\mathcal{A}(t)\in H^\alpha_Q$ for all $t$, \emph{a fortiori} $f_t = Q^{-1}\mathcal{A}(t)\in H^\alpha_C$. Moreover, the dependence of solutions on their initial value is continuous in the $\alpha$-norm.
\end{prop}
\vspace{.2cm}

\noindent
{\em Proof.}  Note that all the operators here have a matrix representation. In particular, the  time-derivative is carried out entry-wise.  Define the constant
\begin{equation}\label{def_c}
  c := 4 \|\mathcal{C}\|_\alpha + 4\sum_{j}\|L_j\|^2_\alpha .
\end{equation}
In view of Proposition \ref{sub_mult}, (\ref{Lindblad_pure}) implies
\begin{equation}\label{est_T2A}
\|\partial_t\mathcal{A}\|_\alpha \leq \frac{c}{2}\|\mathcal{A}\|_\alpha.
\end{equation}
Next, recall that $\mathcal A$ is a matrix operator, and invoke the explicit formula for the $\alpha$-norm to obtain $\partial_t \|\mathcal{A}\|^2_\alpha = 2\Re \langle \partial_t \mathcal{A}| \mathcal{A}\rangle_\alpha$.  Thus,
\[
\partial_t \|\mathcal{A}\|^2_\alpha = 2\Re \langle \partial_t \mathcal{A}| \mathcal{A}\rangle_\alpha \leq  2 \|\partial_t\mathcal{A}\|_\alpha\|\mathcal{A}\|_\alpha \leq  c \,\|\mathcal{A}\|_\alpha^2,
\]
equivalently,
\[
\partial_t \|\mathcal{A}\|_\alpha   \leq  c\, \|\mathcal{A}\|_\alpha.
\]
Therefore
\begin{equation}\label{est_T2}
\|\mathcal{A}(t)\|_\alpha \leq e^{ ct}\|\mathcal{A}(0)\|_\alpha ,
\end{equation}
which completes the proof.
$\Box$
\vspace{.2cm}

It is interesting to consider the case when there is just one Lindblad operator on the right of (\ref{Lindblad_pure}) and, in addition, it has the form:
\begin{equation}\label{L_diagonal}
L = \sum_{n\in\mathbb{Z}}  \lambda_n\, |e_n\rangle\langle e_n|
\mbox{ with }(e_n) \mbox{ an orthonormal basis, and } \lambda_k \neq \lambda_l \mbox{ whenever } k\neq l,
\end{equation}
which is not necessarily in any $H^\alpha_Q$ with $\alpha > 0$.
A direct calculation shows that in this case the solution $\mathcal{A} = [a_{k,l}]$ satisfies
\begin{equation}\label{diag_L}
a_{k,l}(t) = a_{k,l}(0)\exp{(\lambda_k^* \lambda_l -\frac{1}{2}|\lambda_k|^2 -\frac{1}{2}|\lambda_l|^2  )t}.
\end{equation}
Using $(e_n)$ as the distinguished basis to define the Q-transform, the effects of flow (\ref{Lindblad_pure}) can be understood in quite explicit terms. It amounts to a filtering process, namely:
\begin{equation}\label{f_hat_Lind}
\hat{f}_{k,l}(t) = \hat{f}_{k,l}(0)\exp{(\lambda_k^* \lambda_l -\frac{1}{2}|\lambda_k|^2 -\frac{1}{2}|\lambda_l|^2  )t}.
\end{equation}
e.g. if $\lambda_n = c \, n$, where $c$ is a real constant, we have
$
\hat{f}_{k,l}(t) = \hat{f}_{k,l}(0)\exp{(-c^2(k-l)^2t/2)}.
$
Observe that the process suppresses those modes that have a discrepancy in $x$ and $y$ directions; the higher the discrepancy $|k-l|$ the faster the corresponding mode is  suppressed. More generally, we have

\begin{prop} \label{prop_Lind}
Let $\mathcal{A}(t)$ satisfy (\ref{Lindblad_pure}) with $L$ as in (\ref{L_diagonal}), and let $f = Q^{-1}\mathcal{A}(t)$. Assume $ f(0)\in H^\alpha_C$ for an $\alpha \geq 0$. Then $\|f(t)\|_{\alpha}$ is a decreasing function of time, and
 \[
  \lim\limits_{t\rightarrow\infty} \|f(t)\|^2_{\alpha} = \sum\limits_{k\in\mathbb{Z}} (1+2k^2 )^{\alpha} \, |\hat{f}_{k,k}(0)|^2.
  \]
\end{prop}
\vspace{.2cm}

\noindent
{\em Proof.}  The claims follow directly from (\ref{f_hat_Lind}) via the elementary inequality $|\lambda_k|^2+|\lambda_l|^2 \geq 2\Re \, \lambda_k^*\lambda_l$. $\Box$
\vspace{.2cm}

A more general type of dynamic is discussed in the next Subsection.

\subsection{The dissipative flow with the compactly perturbed harmonic oscillator}\label{Main_th}

In this section we make our main observation. Namely, we demonstrate that the quantum  Markovian master equation (\ref{Lindblad_master_dual}) with the Harmonic oscillator type Hamiltonian and operators $L_j$ with finite Sobolev $\alpha$-norms is well posed in $H^\alpha_Q$.

\begin{theorem}\label{thA}
Consider (\ref{Lindblad_master_dual}) wherein all the constituent operators act on the Hilbert space $\mathbb{H}$. Let $(e_n)_{n\in\mathbb{Z}}$  be the distinguished orthonormal basis of $\mathbb{H}$ which determines the Q-transform. Let $\mathcal{C}\in H^\alpha_Q$ ($\alpha \geq 0$) be self-adjoint, and let:
  \[
 \mathcal{H}= \mathcal{H}_0 + \mathcal{C} \mbox{ with } \mathcal{H}_0 = \sum\limits_{n\in \mathbb{Z}} (an+b) |e_n\rangle\langle e_n|\,\, \, (\mbox{for arbitrary } a,b\in \mathbb{R}),
 \]
 and also, let $ L_j \in H^\alpha_Q\,\,  \mbox{ for all }j$.
 Then, the solution of (\ref{Lindblad_master_dual}) satisfies  $\mathcal{A}(t)\in H^\alpha_Q$ (equivalently, $f_t = Q^{-1}\mathcal{A}(t) \in H^\alpha_C$) for all times  $t$, provided $\mathcal{A}(0)\in H^\alpha_Q$ (equivalently, $f_0\in H^\alpha_C$).  Moreover, the dependence of solutions on their initial value is continuous in the $\alpha$-norm.
\end{theorem}
\vspace{.2cm}

\noindent
{\em Proof.} The existence and uniqueness of the solution of (\ref{Lindblad_master_dual}) and its continuous dependence on the initial condition under the stated assumptions all follow directly from Propositions \ref{prop_Harmonic} and \ref{prop_Lind_gen} and the known general results pertaining to perturbation of semigroups via bounded operators, e.g. Chapter IX in \cite{Kato}. Nevertheless, it is interesting to observe the explicit construction of the solution. To this end, consider two separate linear flows:
\[
\partial_t\mathcal{A}_1 = i [\mathcal{H}_0,\mathcal{A}_1]  =: \mathcal{T}_1\mathcal{A}_1
\]
\[
\partial_t\mathcal{A}_2 = i\mathcal{C}\mathcal{A}_2 -i \mathcal{A}_2\mathcal{C} + \sum\limits_j \{L_j^\dagger \mathcal{A}_2 L_j - \frac{1}{2} L_j^\dagger L_j\mathcal{A}_2 - \frac{1}{2} \mathcal{A}_2 L_j^\dagger L_j \}=: \mathcal{T}_2\mathcal{A}_2
\]
We have
$\mathcal{A}_1(t) = \exp{(\mathcal{T}_1t)} \mathcal{A}_1(0) = U(t) \mathcal{A}(t)U(t)^\dagger$ for $U(t) = \exp (i\mathcal{H}_0t)$. By Proposition \ref{prop_Harmonic} we know that the map $\exp{(\mathcal{T}_1t)} $ is an isometry in the $\alpha$-norm.
In addition,  $\mathcal{A}_2(t) = \exp{(\mathcal{T}_2t)} \mathcal{A}_2(0)$.  Applying estimates (\ref{est_T2A}) and (\ref{est_T2}) we obtain
\begin{equation}\label{est_prel}
  \|\mathcal{T}_2\mathcal{A}_2(t)\|_\alpha \leq \frac{c}{2}\exp{(c\, t)}\|\mathcal{A}(0)\|_\alpha,
\end{equation}
 with constant $c$ defined in (\ref{def_c}). The main part of the proof is based on a known approach from the theory of perturbation of semigroups. Namely, the solution $\mathcal{A}(t)$ of (\ref{Lindblad_master_dual}) satisfies the integral equation (see $\S 2$, Chapter IX in \cite{Kato})
\[
\mathcal{A}(t) = U(t) \mathcal{A}(0)U(t)^\dagger + \int\limits_0^t U(t-s) \mathcal{T}_2\,\mathcal{A}(s)\,  U(t-s)^\dagger \, ds.
\]
 Note, again, that all the operators here have a matrix representation. In particular, the time-integral  is carried out entry-wise.
The integral identity implies
\[
\|\mathcal{A}(t)\|_\alpha \leq \|U(t) \mathcal{A}(0)U(t)^\dagger\|_\alpha + \|\int\limits_0^t U(t-s) \mathcal{T}_2\,\mathcal{A}(s)\,  U(t-s)^\dagger \, ds\,\|_\alpha .
\]
 Since $U(t)$ is diagonal by (\ref{invariance}) we have $$\|U(t) \mathcal{A}(0)U(t)^\dagger\|_\alpha = \|\mathcal{A}(0)\|_\alpha .$$
Next, for convenience, we let $\left(M\right)_{k,l} $ denote the $(k,l)$ entry of the matrix $M$. Thus,
\[
\begin{array}{lr}
\|\int\limits_0^t U(t-s) \mathcal{T}_2\,\mathcal{A}(s)\,  U(t-s)^\dagger \, ds\,\|_\alpha ^2 =& \\
 & \\
 \sum\limits_{k,l} (1+k^2+l^2)^\alpha \left|\int\limits_0^t \left(U(t-s) \mathcal{T}_2\,\mathcal{A}(s)\,  U(t-s)^\dagger\right)_{k,l} \, ds\right|^2 &  \\
   &  \\
  \leq t \int\limits_{0}^{t}\, \sum\limits_{k,l} (1+k^2+l^2)^\alpha \left|\left(U(t-s) \mathcal{T}_2\,\mathcal{A}(s)\,  U(t-s)^\dagger\right)_{k,l}\right|^2\, ds &   \\
   & \\
  =t\, \int\limits_{0}^{t}\, \|\mathcal{T}_2\,\mathcal{A}(s)\|_\alpha^2\, ds    &
  \mbox{( by (\ref{invariance}) )} \\
  & \\
 \leq t\,\int\limits_{0}^{t}\,\frac{c^2}{4} e^{2cs}\,\|\mathcal{A}(0)\|_\alpha^2\, ds & \mbox{( by (\ref{est_prel}) )}\\
 & \\
  = \frac{1}{8}t\, c\,(e^{2ct}-1) \, \|\mathcal{A}(0)\|_\alpha^2. &
\end{array}
\]
In summary, we obtain
\begin{equation}\label{estimate_full}
  \|\mathcal{A}(t)\|_\alpha \leq \left(1+ \sqrt{c \,t\, (e^{2ct}-1)}/(2\sqrt{2})\right)\,\, \|\mathcal{A}(0)\|_\alpha.
\end{equation}
Thus, a solution whose initial value is in $H^\alpha_Q$ remains in this space for all times $t>0$. The estimate also implies that the dependence of solutions on the initial condition is continuous in this space. $\Box$
\vspace{.2cm}

\noindent
\textbf{Remark 1.} Theorem \ref{thA} is formulated in the Heisenberg picture. A homologous statement in easily obtained in the dual setting (the  Schr\"{o}dinger picture), in which the mixed state, say, $\rho=\rho(t)$ is undergoing evolution. The corresponding equation of motion is just slightly modified, \cite{Alicki}. It is easily seen that one may replace $\mathcal{A}(t)$ by $\rho(t)$ in all the statements in Theorem \ref{thA} and its proof. The result, in either picture, admits an interesting interpretation. Namely, the Sobolev norms capture the regularity  of a state (or observable), as defined via the Q-transform, and the theorem implies that they cannot de-regularize too fast during quantum evolution.  This sheds some light on the nature of some challenges that are encountered in the course of numerical simulation of quantum systems. Indeed, classical simulation of quantum processes is quite efficient for small system sizes but becomes unfeasible very quickly as the system size increases. The main barriers to efficiency of larger scale simulations stem from memory requirements, e.g. see \cite{Sowa_Zagoskin}. At the same time, data sets such as smooth functions are well compressible, i.e. they may be represented in a classical digital environment with good accuracy, requiring relatively little space in the computer memory.  Such representations are facilitated via the so-called fast transforms. Generally, the smoother the function the more compressible it is. Let us consider, hypothetically, that  an evolving quantum observable is such that its classical alter ego, obtained via the Q-transform, remains smooth in the time interval of interest. Since it is possible to store information about the observable quite efficiently the memory limitations would not be a barrier to classical simulation of the underlying quantum dynamic. Now,  Theorem \ref{thA} suggests that the considered scenario is not unreasonable.  Indeed, in the finite-dimensional setting, estimate (\ref{estimate_full}) --- or the slightly sharper (\ref{est_T2}), which also applies --- implies that an observable (or state) cannot de-regularize too quickly during evolution.  In fact, numerical experiments suggest that in many special systems the observable fully retains its regularity during quantum evolution. Of course, compressibility of the dynamic variable by itself is not sufficient to conclude that the entire evolutionary process can be efficiently simulated. Indeed, there remains the complementary problem of whether or not the rule of evolution can also be implemented via efficient algorithms. The latter problem is of a different nature and will not be addressed here.
\vspace{.2cm}

\noindent
\textbf{Remark 2.} It is interesting to mention that in light of the Sobolev embedding theorem all functions in $H^\alpha_C$ are continuous whenever $\alpha > 1$, thus the nonlocal dynamics in the classical variable $f_t$ is well-posed in the space of continuous functions.
Our interest in the general Sobolev space setting stems from the method of proof of Theorem \ref{thA}, as well as the fact that natural images are typically represented by functions with discontinuities. A study of an application of this type of dynamics to image enhancement is reported in \cite{Sowa33}, which exploits efficient numerical algorithms developed in \cite{Sowa_Zagoskin}.

\section{Generalized Q-transforms}

In this section we introduce analogons of the Q-transform obtained via certain special nonorthogonal bases in the Hilbert spaces $L_2(\mathbb{R}/\mathbb{Z})$ and $L_2(\mathbb{T})$. The first main outcome is that there exist a plethora of generalized Q-transforms, i.e. a variety of ways in which a periodic distribution can be encoded into a quantum observable. The second is a discussion of the phenomenon of broadband redundancy (see Theorem \ref{thB}). In essence, it shows that any observable may be seen as an average of a plethora of other observables that are less regular in comparison with the original one.

\subsection{The generalized Q-transform related to the Dirichlet ring}
We begin by a brief description of the nonorthogonal transforms. Let $L_2(\mathbb{R}/\mathbb{Z}) = \mathbb{H}_a\oplus \mathbb{H}_h$ be a direct sum decomposition into the non-positive frequency and positive frequency subspaces. We consider a family of functions in $\mathbb{H}_h$ defined by a square-summable sequence of complex numbers $(a_l)_{l=1}^{\infty}$ in the following way
\begin{equation}\label{deff}
\varphi_m(t) = \sum\limits_{l=1}^\infty a_l\,e^{2\pi imlt},
\quad m =1,2,3,\ldots
\end{equation}
Observe the scaling $\varphi_m(t) = \varphi_1(mt)$. Also, all $\varphi_m$ have identical length (i.e. $L_2$-norm) $||\varphi_m|| = (\sum |a_l|^2)^{1/2} < \infty$. Furthermore, let $D: \mathbb{H}_h\rightarrow \mathbb{H}_h $ be the change of basis map, i.e. $D e_m = \varphi_m$ where $e_m(t) = \exp{(2\pi im t)}$. It is known (see Theorem 2 in \cite{sowa21}) that if $(a_l)$ together and its Dirichlet inverse $(b_l)$ are both absolutely summable, then $D$ \emph{is a linear homeomorphism of} $\mathbb{H}_h$. Recall that the Dirichlet inverse of $(a_l)$ is defined recursively via, \cite{Apostol2}:
\[
b_1 = 1/a_1, \qquad b_n = -\frac{1}{a_1} \sum\limits_{d|n, d>1} a_d\,b_{n/d}\quad n = 2,3, \ldots ,
\]
where $\sum_{d|n, d>1}$ denotes the sum over all divisors $d$ of $n$ except $d=1$. Since all vectors $\varphi_m$ have the same length, $D$ being a homeomorphism means that, by definition, $(\varphi_m)_{m\in\mathbb{N}}$ is a Riesz basis in $\mathbb{H}_h$, \cite{Gohberg}. Its dual is the basis $(\chi_n)$, given by
\[
\chi_n = \sum\limits_{d|n}b_d^* e_{\frac{n}{d}}, \mbox{ so that } \langle \chi_n|\varphi_m\rangle = \delta_{m,n}
\]
It is known that the dual of a Riesz basis is also a Riesz basis.
Let us set $\varphi_0 = 1 = \chi_0$ (the constant function), and $\varphi_{-m} = \varphi_m^*$, $\chi_{-m} = \chi_m^*$, so that, trivially, the extended family $(\varphi_m)_{m\in\mathbb{Z}}$, and its dual $(\chi_m)_{m\in\mathbb{Z}}$, are both Riesz bases in $L_2(\mathbb{R}/\mathbb{Z})$. Also, extending $D$ by setting $D e_{-m} = \varphi_{-m} $ makes it a homeomorphism of $L_2(\mathbb{R}/\mathbb{Z})$.
It is demonstrated in \cite{sowa21} that $D, D^{-1}:\mathbb{H}_h \rightarrow \mathbb{H}_h$ are matrix operators (with a particular structure related to the Dirichlet ring) and, obviously so are their extensions to $L_2(\mathbb{R}/\mathbb{Z})$. Namely, retaining the same notation $D$ for the operator extended from $\mathbb{H}_h$ to the entire $L_2(\mathbb{R}/\mathbb{Z})\equiv \ell_2(\mathbb{Z})$, we have
\begin{equation}\label{Matrix_formal}
D =\begin{array}{lllllllllllllllllll}
 &\vdots &\vdots &\vdots &\vdots &\vdots  &\vdots &\vdots &\vdots &\vdots &\vdots &\vdots &\vdots &\vdots  &\vdots &\vdots &\vdots &\\
 \ldots &\cdot&a_1^*&\cdot&\cdot&a_2^*&a_3^*&a_6^*&\cdot&\cdot &\cdot &\cdot &\cdot &\cdot &\cdot &\cdot &\cdot &\ldots\\
 \ldots &\cdot&\cdot&a_1^*&\cdot&\cdot&\cdot&a_5^*&\cdot&\cdot &\cdot &\cdot &\cdot &\cdot &\cdot &\cdot &\cdot &\ldots\\
 \ldots &\cdot&\cdot&\cdot&a_1^*&\cdot&a_2^*&a_4^*&\cdot&\cdot &\cdot &\cdot &\cdot &\cdot &\cdot &\cdot &\cdot &\ldots\\
 \ldots &\cdot&\cdot&\cdot&\cdot&a_1^*&\cdot&a_3^*&\cdot&\cdot &\cdot &\cdot &\cdot &\cdot &\cdot &\cdot &\cdot &\ldots\\
 \ldots &\cdot&\cdot&\cdot&\cdot&\cdot&a_1^*&a_2^*&\cdot&\cdot &\cdot &\cdot &\cdot &\cdot &\cdot &\cdot &\cdot &\ldots\\
 \ldots &\cdot&\cdot&\cdot&\cdot&\cdot&\cdot&a_1^*&\cdot&\cdot &\cdot &\cdot &\cdot &\cdot &\cdot &\cdot &\cdot &\ldots\\
 \ldots &\cdot&\cdot&\cdot&\cdot&\cdot&\cdot&\cdot&1&\cdot &\cdot &\cdot &\cdot &\cdot &\cdot &\cdot &\cdot &\ldots\\
 \ldots &\cdot&\cdot&\cdot&\cdot&\cdot&\cdot&\cdot&\cdot&a_1 &\cdot &\cdot &\cdot &\cdot &\cdot &\cdot &\cdot &\ldots\\
 \ldots &\cdot&\cdot&\cdot&\cdot&\cdot&\cdot&\cdot&\cdot &a_2 &a_1 &\cdot &\cdot &\cdot &\cdot &\cdot &\cdot &\ldots\\
  \ldots&\cdot&\cdot&\cdot&\cdot&\cdot&\cdot&\cdot&\cdot &a_3 &\cdot &a_1 &\cdot &\cdot &\cdot &\cdot &\cdot &\ldots\\
  \ldots&\cdot&\cdot&\cdot&\cdot&\cdot&\cdot&\cdot&\cdot &a_4 &a_2 &\cdot &a_1 &\cdot &\cdot &\cdot &\cdot &\ldots\\
 \ldots &\cdot&\cdot&\cdot&\cdot&\cdot&\cdot&\cdot&\cdot &a_5 &\cdot &\cdot &\cdot &a_1 &\cdot &\cdot &\cdot &\ldots\\
  \ldots&\cdot&\cdot&\cdot&\cdot&\cdot&\cdot&\cdot&\cdot &a_6 &a_3 &a_2 &\cdot &\cdot &a_1 &\cdot &\cdot &\ldots\\
 &\vdots &\vdots &\vdots &\vdots &\vdots  &\vdots &\vdots &\vdots &\vdots &\vdots &\vdots &\vdots &\vdots  &\vdots &\vdots &\vdots &
 \end{array}
\end{equation}
where we have substituted dots for zeros to de-clutter the appearance of the matrix. Remarkably, the matrix for $D^{-1}$ has the same structure as that of $D$ and is obtained from the latter by substituting coefficients $b_k$ for $a_k$, \cite{sowa21}. Observe that the entries of such matrices contain an infinite number of copies of each element of the sequence $(a_k)$. Hence, neither $D$ nor its inverse can be Hilbert-Schmidt or, \emph{a fortiori}, in $H^\alpha_Q$  as long as $\alpha \geq 0$. However, we have a convenient estimate for the operator norm of $D$.
Indeed, for a matrix operator $A$ on $\ell_2 = \ell_2(\mathbb{Z})$, let us define
\[
C_1 = \sup\limits_l \sum\limits_k |A_{kl}|, \quad C_\infty = \sup\limits_k \sum\limits_l |A_{kl}|.
\]
It is well known (see Chapter III, \cite{Kato}) that if $C_1$ and $C_\infty$ are both finite, the operator norm of $A$ is bounded and satisfies
\[
  \|A\| := \|A\|_{\ell_2\rightarrow\ell_2} \leq (C_\infty)^{\frac{1}{2}}\, (C_1)^{\frac{1}{2}}
\]
Next, observe that for $A=D$ as in (\ref{Matrix_formal}) we have $C_1 = \max\{1,\sum_{n=1}^\infty |a_n|\}$, while $C_\infty \leq C_1$. Hence,
\begin{equation}\label{D_oper_norm}
  \|D\| \leq \max\{1, \sum\limits_{n=1}^\infty |a_n|\}.
\end{equation}
(Note that this provides useful information only if the sequence $(a_l)$ is indeed summable.)

Next, we fix the tensor product basis in $L_2(\mathbb{T})$, which consists of functions $\varphi_{m,n}(x,y) = \varphi_m(x)\varphi_n(y)$ for $m,n \in\mathbb{Z}$. Note that $D^{\otimes 2}$ is a homeomorphism of $L_2(\mathbb{T})$. In summary a sequence $(a_l)$ with suitable properties determines a basis in the space of square integrable functions on the torus. Moreover, the construction of the basis $(\varphi_{k,l})$ ensures the same type of symmetry as that of the Fourier transform. Namely, for $ f\in L_2(\mathbb{T})$, the $\varphi_{k,l}$-coefficients of $f$ are defined via
\[
z_{k,l}   =  \int\limits_0^{1}\int\limits_0^{1} f(x,y) \chi_{-k,-l}(x,y) \, dxdy =  \int\limits_0^{1}dx\,\chi_{k}^*(x)\int\limits_0^{1}dy f(x,y) \chi_{l}^*(y) .
\]
Thus, if $f$ is real-valued, then the matrix $Z = [z_{k,l}]_{(k,l)\in\mathbb{Z}^2} $ satisfies the symmetry condition (\ref{point-sym}).
Note that
\begin{equation}\label{D-transform}
Z =  D^{-1}\hat{f} \,(D^{-1})^T \qquad \mbox{ where } \hat{f} = [\hat{f}_{k,l}]_{(k,l)\in\mathbb{Z}^2},
\end{equation}
and the superscript $^T$ denotes matrix transposition.
It is useful to describe $z_{k,l}$ directly.
 To describe the general case efficiently we adopt the following conventions. First, for $k>0$ we denote $b_{-k} = b_k^*$; also, $b_0=1$. Secondly, when $k<0$, $d|k$ means $d$ is a (positive) divisor of $|k|$; also $d|0$ means $d=1$. Finally, we let $\sgn(k)$ denote the sign of $k$ with the proviso $\sgn(0) = 0$. Then, we have
\begin{equation}\label{D-transform-indices_neg}
z_{k,l} = \sum\limits_{d|k}\sum\limits_{r|l}b_{\frac{k}{d}} \,b_{\frac{l}{r}}\, \hat{f}_{\sgn(k) d, \sgn(l)r} =
\sum\limits_{d|k}\sum\limits_{r|l}b_{\sgn(k)d} \,b_{\sgn(l)r}\, \hat{f}_{\frac{k}{d},\frac{l}{r}} .
\end{equation}
For a periodic distribution $\omega$, its D-transform is defined to be
\begin{equation}\label{D-transform-dist}
Z = D^{-1}\hat{\omega}\,(D^{-1})^T\qquad \mbox{ where } \hat{\omega} = [\hat{\omega}_{k,l}]_{(k,l)\in\mathbb{Z}^2}.
\end{equation}
 Of course, this is a formal definition and the coefficients may or may not be well-defined, depending on the regularity properties of the functions $(\varphi_{m,n})$, which stem from properties of the sequence $(a_l)$. In the same manner, we define a generalized Q-transform, denoted $Q^D$, by setting
 \begin{equation}\label{QD}
Q^D\omega = S[D^{-1}\hat{\omega} \,(D^{-1})^T],
\end{equation}
where $S$ is as in Subsection \ref{subsection-define-S}. The following result is almost immediate:

\begin{prop} \label{prop_D_HS}
Assume that $D$ is a homeomorphism of $L_2(\mathbb{R}/\mathbb{Z})$. Then, for every $f\in L_2(\mathbb{T})=H^0_C$, $Q^D f$ is a Hilbert-Schmidt operator. Moreover,
\[
\|Q^D f\|_0 \leq \|D^{-1}\|^2\, \|f\|_0.
\]
\end{prop}
\vspace{.2cm}

\noindent
{\em Proof.}  
 If$f\in H^0_C$, then the matrix $\hat{f}$ represents a Hilbert-Schmidt operator. It is well known, see e.g. \cite{Shubin} that the subspace of Hilbert -Schmidt operators is a two-sided ideal in the space of all bounded operators, hence $D^{-1}\hat{f} \,(D^{-1})^T\in H^0_C$. Moreover, the estimate of the Hilbert-Schmidt norm of $Q^Df$ follows from (\ref{QD}). Indeed, it is known that for bounded $A,B$ and Hilbert-Schmidt $C$ one has
 $\|AC\|_0\leq \|A\|\, \|C\|_0$ as well as  $\|CB\|_0\leq \|B\|\, \|C\|_0$. Therefore,
\[
\|Q^D f\|_0 = \|D^{-1}\hat{f} \,(D^{-1})^T\|_0 \leq \|D^{-1}\|\, \|\hat{f}\|_0 \,\|(D^{-1})^T\| =  \|D^{-1}\|^2\, \|f\|_0,
\]
as claimed. $\Box$
\vspace{.1cm}

\noindent
\textbf{Remark 1.} As is the case for the Q-transform the $Q^D$-transform is also naturally extended to complex-valued distributions via the relation $Q^D(f+ig) := Q^Df +iQ^Dg$.
\vspace{.1cm}

\noindent
\textbf{Remark 2.} More properties of the action of matrices (\ref{Matrix_formal}) in  $\ell_2$ and other select sequence spaces are described  in \cite{sowa21} and \cite{Sowa_Applcat_Math}.

\subsection{The periodized zeta function, and the broadband redundancy}

It was observed in \cite{Sowa34} that a pure tone, say $\sin x$, can be well approximated by averages of certain special broadband signals (i.e. periodic functions whose Fourier coefficients decay relatively slowly). This is due to a cancellation of higher frequency components. In order to properly describe this phenomenon, we need to discuss the periodized Riemann zeta function, for details see \cite{Apostol2},
\begin{equation} \label{periodized_def}
F_s(t) = F(s, t) = \sum\limits_{k=1}^{\infty} \frac{e^{2\pi i kt}}{k^s},\quad t \in \mathbb{R}/\mathbb{Z}, s=\sigma + i\tau \in \mathbb{C}.
\end{equation}
The series converges absolutely whenever $\sigma > 1$ and, for a fixed $t$, defines an analytic function of $s$, say, $F_t(s)$. $F_t(s)$ is extended by analytic continuation to the entire complex plane. It is holomorphic everywhere except for the point $s=1$ where $F_t(s)$  has a simple pole.  The Riemann zeta function is the special case $\zeta(s) = F_1(s)$. Assuming the Riemann Hypothesis (RH), let $s_n = 1/2 + i\tau_n$ be the sequence of zeros of $\zeta$ in the upper-right half plane\footnote{The famous RH is the statement: $\sigma_n = 1/2$ for all $n$. In what follows we use a broadly known result from \cite{Fujii2}, which relies upon the RH. However, an unconditional proof of this result was reported in \cite{Steuding}.}, arranged in the order of nondecreasing imaginary parts, so that $\tau_n\leq \tau_{n+1}$. It has been known for quite some time now, \cite{Landau}, that for every real number $\alpha \neq 0$ the sequence $(\alpha \, \tau_n)_{n=1}^{\infty}$ is uniformly distributed modulo one. For this reason, as will be clarified in what follows, it is enticing to consider bases of the form (\ref{deff}) when, specifically,
\begin{equation}\label{def_spec_phi}
  \varphi_m(t) = \varphi_{m,[\sigma + i\tau_n]}(t) = F(\sigma + i\tau_n, mt), \quad m =1,2,3,\ldots,
\end{equation}
where the subscript $[\sigma + i\tau_n]$ indicates dependence on the choice of the Riemann zero ordinate $\tau_n$ and arbitrary $\sigma >0$.
  We will denote the corresponding maps $D$  by $D_{[\sigma + i\tau_n]}$. It is in every instance clear from the context whether $D_{[\sigma + i\tau_n]}$ is meant as an operator in $L_2(\mathbb{R}/\mathbb{Z})$, or equivalently a matrix operator in $\ell_2(\mathbb{Z})\equiv L_2(\mathbb{R}/\mathbb{Z})$; hence, for simplicity, we use the same symbol in all cases.
 In particular, for a fixed $\sigma$ and $n$, we have\footnote{The statement (\ref{special_a_n}) about the Fourier coefficients of the periodized zeta function is trivial for $\sigma >1$, but requires a nontrivial argument when $1\geq \sigma>0$, see \cite{Sowa34}. In this article we will only consider the former case.}
\begin{equation}\label{special_a_n}
 a_k = k^{-\sigma - i\tau_n }, \quad k =1,2,3,\ldots \mbox{ for all } \sigma >0.
 \end{equation}
 It is immediate that the coefficients $b_n$ assume of the form
\begin{equation}\label{special_b_n}
 b_k = \mu(k) k^{-\sigma - i\tau_n }, \quad k = 1, 2,3,\ldots \mbox{ for all } \sigma >0,
 \end{equation}
where $\mu$ is the M\"{o}bius function, i.e. $\mu(1) =1$, and for $k>1$: $\mu(k) = 0$ if $k$ is divisible by a square, $\mu(k) = 1$ if $k$ is a product of an even number of mutually distinct primes, and $\mu(k) =-1$ if $k$ is a product of an odd number of mutually distinct primes. Indeed, it is trivially seen that $(b_n)$ is the Dirichlet inverse of $(a_n)$.
In light of this we have
\begin{equation}\label{homeo}
  \mbox{ If } \sigma > 1, \mbox{ then } D_{[\sigma + i\tau_n]}\mbox{ is a homeomorphism } (\mbox{of } \ell_2(\mathbb{Z})\equiv L_2(\mathbb{R}/\mathbb{Z})).
\end{equation}
Indeed, $\sigma >1$ implies that both sequences $a_n$ and $b_n$ are absolutely summable.
Note also that under the same conditions operation $f \mapsto Z$ defined in (\ref{D-transform}) gives a homeomorphic map between $L_2(\mathbb{T})$ and $\ell_2(\mathbb{Z}^2)$.

Next, let $N(T)$ denote the number of $\tau_n$'s contained in the interval $(0,T]$. It is a direct consequence of the result of Fujii, \cite{Fujii2}, that whenever $\sigma >0$ there is a positive constant $C$ such that
\begin{equation}\label{convergence}
 \left|\frac{1}{N(T)}\sum\limits_{\tau_n\leq T} k^{-\sigma - i\tau_n }\right| \leq C  \frac{k^{1/2-\sigma}\log k}{\log T}\quad \mbox{ for all }    k \geq 2 \mbox{ and } T \mbox{ sufficiently large. }
\end{equation}
In what follows we exploit this fact to describe the redundancy inherent in the homeomorphisms $D_{[\sigma + i\tau_n]}$. and, by the same token, in the corresponding generalized Q-transforms.
The following helps develop a method of proof that will be used again in Theorem \ref{thB}.
\begin{theorem} \label{theorem_special_D}
Let $f\in  H^\alpha(\mathbb{R}/\mathbb{Z})$, i.e. $\|f\|_\alpha^2 =\sum_{k\in\mathbb{Z}}  (1+k^2)^\alpha |\hat{f}_k|^2 <\infty$, ($\alpha\geq 0$).
Then,
\[
\left\|\frac{1}{N(T)}\sum\limits_{\tau_n\leq T} D_{[\sigma + i\tau_n]} \hat{f} - \hat{f}\right\|_{\ell_2} \longrightarrow 0 \,\mbox{ as } T \rightarrow \infty, \mbox{ provided } \alpha >1/2 \mbox{ and } \sigma > \alpha+1,
\]
and the rate of convergence depends on $f$ only via the value of its $\alpha$-norm.
\end{theorem}
\vspace{.2cm}

\noindent
{\em Proof.} Since $\sigma >3/2$ by assumption $(\varphi_{k, [\sigma + i\tau_n]})$ is a basis in $L_2(\mathbb{R}/\mathbb{Z})$ (see the argument following (\ref{homeo})). Let $(\chi_{k, [\sigma + i\tau_n]})$ be the dual basis. Consider the coefficients
\[
z_k =  \frac{1}{N(T)}\sum\limits_{\tau_n\leq T} (D_{[\sigma + i\tau_n]} \hat{f}) (k) =  \frac{1}{N(T)}\sum\limits_{\tau_n\leq T}\, \int\limits_0^{1} f(x) \chi_{-k, [\sigma + i\tau_n]}(x) \, dx.
\]
It follows from (\ref{special_b_n}) that, for $k\geq 1$,
\begin{equation}\label{z}
z_k = \sum\limits_{d|k} \mu(d) \frac{1}{N(T)}\sum\limits_{\tau_n\leq T} d^{-(\sigma+i\tau_n)} \hat{f}_{k/d}= \hat{f}_k + \sum\limits_{d|k, d>1} \mu(d) \frac{1}{N(T)}\sum\limits_{\tau_n\leq T} d^{-(\sigma+i\tau_n)} \hat{f}_{k/d}.
\end{equation}
For $d>1$ let us denote
\begin{equation}\label{def_cd}
c_d(T) = \left|\frac{1}{N(T)}\sum\limits_{\tau_n\leq T} d^{-(\sigma+i\tau_n)} \right|,
\end{equation}
and observe that by (\ref{convergence}) there exists $C>0$ independent of $d$, such that for any $\epsilon >0$, we have
\begin{equation}\label{ineq_cd}
c_d(T) < \frac{C}{\log T} d^{1/2 -\sigma}\log d
\end{equation}
Next, fix $\alpha \geq 0$. From (\ref{z}) we obtain
\[
\begin{array}{rll}
  |z_k -\hat{f}_k|^2 \leq  & \left(\sum\limits_{d|k, d>1}  c_d(T) |\hat{f}_{k/d}|\right)^2 & =\left(\sum\limits_{d|k, d>1}  c_d(T) \,\left(\frac{k}{d}\right)^{-\alpha}\,\left(\frac{k}{d}\right)^\alpha\, |\hat{f}_{k/d}|\right)^2
  \end{array}
  \]
and, by Cauchy-Schwartz,
  \[
  \begin{array}{rll}
  & & \\
  \leq & \sum\limits_{d|k, d>1}  c_d(T)^2 \left(\frac{k}{d}\right)^{-2\alpha}
  \sum\limits_{d|k, d>1}\left(\frac{k}{d}\right)^{2\alpha}|\hat{f}_{k/d}|^2 & \leq \|f\|_\alpha^2
  \sum\limits_{d|k, d>1}  c_d(T)^2 \left(\frac{k}{d}\right)^{-2\alpha} \\
  & & \\
  & \leq \|f\|_\alpha^2  \left(\frac{C}{\log T}\right)^2 k^{-2\alpha}\, \sum\limits_{d|k, d>1} d^{1 -2(\sigma-\alpha)} (\log d )^2&
  \leq  \|f\|_\alpha^2  \left(\frac{C}{\log T}\right)^2 k^{-2\alpha}\, \zeta''(2(\sigma-\alpha) -1) \\
  & & \\
  & &  \mbox{ provided } \sigma > \alpha+1.
\end{array}
\]
In the last step we have used the fact that the series $\sum_n (\log n)^2 n^{-s} $ converges to $\zeta''(s)$, provided $\sigma = \Re s > 1$.
The case of negative coefficients $k$ is handled analogously, taking positive divisors $d$ of $k$ where appropriate, inserting $|k/d|^\alpha$, and replacing $\sigma + i\tau_n$ with the complex conjugate. This leads to an estimate:
\[
|z_k -\hat{f}_k|^2 \leq  \|f\|_\alpha^2  \left(\frac{C}{\log T}\right)^2 |k|^{-2\alpha}\, \zeta''(2(\sigma-\alpha) -1)  \mbox { for } k<0.
\]
Also note $z_0 =\hat{f}_0$.
It follows that
\[
\sum_k|z_k - \hat{f}_k|^2 \leq \|f\|_\alpha^2  \left(\frac{C}{\log T}\right)^2 \, \zeta''(2(\sigma-\alpha) -1)\sum\limits_{k\neq 0} |k|^{-2\alpha}\rightarrow 0
 \]
 as $T\rightarrow \infty$, provided $\alpha>1/2$. This completes the proof. $\Box$
\vspace{.2cm}

\noindent
\textbf{Remark.} We point out that Theorem \ref{theorem_special_D} is somewhat reminiscent of the properties of the classical Dirichlet and Fej\'{e}r kernels, wherein the function $f$ is represented as a limit of trigonometric polynomials. However, in here a function, possibly a very smooth one such as a trigonometric polynomial, is represented as a limit of functions rich in high-frequency content. In other words, smoothness arises out of cancellation of noise, which is somewhat reminiscent of the Central Limit Theorem\footnote{The reader may wish to compare this type of pseudorandomness to the pseudorandom aspects of the Dirichlet series, e.g. as emphasized in \cite{Montgomery}.}. This is the essence of the \emph{broadband redundancy}. In the closing of this article we will describe this phenomenon in the context of generalized Q-transforms.

We let $Q^{[\sigma+i\tau_n]}$ denote the generalized Q-transform corresponding to $D_{[\sigma+i\tau_n]}$ via (\ref{QD}). We have the following

\begin{theorem} \label{thB} Let $\alpha>1/2$ and $\sigma > \alpha +1$. For $f\in  H^{2\alpha}(\mathbb{T})$ we have
\begin{equation}\label{Ba}
\left\|\frac{1}{N(T)}\sum\limits_{\tau_n\leq T} D^{-1}_{[\sigma + i\tau_n]} \hat{f}\, (D^{-1}_{[\sigma + i\tau_n]} )^T - \hat{f}\,\right\|_{\ell_2(\mathbb{Z}\times \mathbb{Z})} \longrightarrow 0 \,\mbox{ as } T \rightarrow \infty,
\end{equation}
and also
\begin{equation}\label{Bb}
\left\|\frac{1}{N(T)}\sum\limits_{\tau_n\leq T} Q^{[\sigma + i\tau_n]} f- Qf\,\right\|_{0} \longrightarrow 0 \,\mbox{ as } T \rightarrow \infty.
\end{equation}
In both cases the rate of convergence depends  on $f$ only via its $2\alpha$-norm.

\end{theorem}
\vspace{.2cm}

\noindent
{\em Proof.} Since $S$ in (\ref{QD}) is an isometry statements (\ref{Ba}) and (\ref{Bb}) are equivalent. We will demonstrate (\ref{Ba}) applying a strategy akin to that developed in the proof of Theorem \ref{theorem_special_D}. Let
 \[
z_{k,l} =  \frac{1}{N(T)}\sum\limits_{\tau_n\leq T} \left(D^{-1}_{[\sigma + i\tau_n]} \hat{f}\, (D^{-1}_{[\sigma + i\tau_n]} )^T \right) (k,l) .
\]
First, we focus on the case when both $k$ and $l$ are positive. By assumption $\sigma >3/2$ and hence $D^{-1}_{[\sigma + i\tau_n]} $  are homeomorphisms. Recall the matrices are populated with coefficients $b_n = \mu(n) n^{-(\sigma + i\tau_n)}$ and their complex conjugates according to the pattern (\ref{Matrix_formal}). Thus, we obtain
\[
z_{k,l} - \hat{f}_{k,l} = \sum\limits_{d|k,r|l,rd>1} \mu(d)\mu(r) \frac{1}{N(T)}\sum\limits_{\tau_n\leq T} (rd)^{-(\sigma + i\tau_n)}\left|\hat{f}_{\frac{k}{d}, \frac{l}{r}}\right|
\]
Next, we fix $\alpha\geq 0$. Using (\ref{def_cd}) we obtain
\[
|z_{k,l} - \hat{f}_{k,l}| \leq \sum\limits_{d|k,r|l,rd>1} c_{rd}(T) \left|\hat{f}_{\frac{k}{d}, \frac{l}{r}}\right| =
\sum\limits_{d|k,r|l,rd>1} c_{rd}(T) \left(\frac{kl}{rd}\right)^{-\alpha} \left(\frac{kl}{rd}\right)^{\alpha}
 \left|\hat{f}_{\frac{k}{d}, \frac{l}{r}}\right| ,
\]
and by Cauchy-Schwartz
\[
|z_{k,l} - \hat{f}_{k,l}|^2 \leq
\sum\limits_{d|k,r|l,rd>1} c_{rd}(T)^2 \left(\frac{kl}{rd}\right)^{-2\alpha} \sum\limits_{d|k,r|l,rd>1} \left(\frac{kl}{rd}\right)^{2\alpha}
 \left|\hat{f}_{\frac{k}{d}, \frac{l}{r}}\right|^2
\]
The trivial inequality $(kl/(rd))^2\leq (1+(k/d)^2+(l/r)^2)^2$ implies that the second sum is bounded by $\|f\|_{2\alpha}^2$. Also, applying (\ref{ineq_cd}), we obtain
\[
\begin{array}{ll}
  \sum\limits_{d|k,r|l,rd>1} c_{rd}(T)^2 \left(\frac{1}{rd}\right)^{-2\alpha} & \leq \left(\frac{C}{\log T} \right)^2 \sum\limits_{r|l} r^{1 - 2(\sigma-\alpha)}(\log r)^2 \,\sum\limits_{d|k} d^{1 - 2(\sigma-\alpha)}(\log d)^2  \\
   &  \\
   & \leq \left(\frac{C}{\log T}\, \zeta''(2(\sigma-\alpha)-1)\right)^2
\end{array}
\]
provided $\sigma > \alpha +1$.
The other quadrants in the $(k,l)$-grid are handled similarly when both $k$ and $l$ are nonzero. Finally, when one of the indices is zero, say, $l=0$, $k>0$, the expression $(kl)^{-2\alpha}$ in the estimates as above is simply replaced by $k^{-2\alpha}$ and the sum $\sum_{r|l}$ is simply erased, etc. In summary, we obtain
\[
\sum\limits_{(k,l)\in\mathbb{Z}^2}|z_{k,l} - \hat{f}_{k,l}|^2 \leq
\left(\frac{C}{\log T} \zeta''(2(\sigma-\alpha)-1)\right)^2\|f\|_{2\alpha}^2 \sum\limits_{k\neq 0} |k|^{-2\alpha} \sum\limits_{l\neq 0} |l|^{-2\alpha}.
\]
The right hand side is finite whenever $\alpha >1/2$.
This completes the proof.  $\Box$

\section*{Summary}
We have observed that classical 2D spatial data may be encoded into a quantum observable and vice versa. This has made possible a discussion of the problem of well-posedness for the evolution of observables in an open quantum system.  It was demonstrated  that under some conditions the Sobolev regularity of quantum observables, as defined in the present work, is preserved during their evolution in a quantum channel. This points at ways of addressing the problem of \emph{compressibility} of quantum information. In addition, we have examined the phenomenon of \emph{broadband redundancy} making a start toward an assessment of its status in the quantum reality.

Readers familiar with the Wigner transform, \cite{Gosson}, may wonder if the Q-transform is in any way related to it. A simple inspection makes clear that the two transforms are essentially different and to an extent complimentary. The first difference is that in the present formulation the Q-transform deals with bi-periodic functions while the Wigner transform deals with functions that are defined in the plane and have some localization property. However, this by itself is rather superficial as, indeed, the Q-transform may be extended to the planar setting\footnote{Such an extension will be discussed in future work.}.  More significantly, the two transforms engage the symmetry of self-adjointness differently and, in addition, rely upon different Fourier transforms (one-dimensional in the case of the Wigner transform versus two-dimensional in the case of the Q-transform).

\section*{Acknowledgments} I am immensely grateful to Alexandre Zagoskin and Paul Babyn for inspiring conversations. Also, I gratefully acknowledge the 2016 Compute Canada Resource Allocation award.


\end{document}